\documentclass{llncs}
\usepackage{hyperref}
\usepackage{array}
\usepackage{graphicx}
\usepackage{tcolorbox}
\usepackage{tikz}
\usetikzlibrary{shapes.geometric, arrows}
\usepackage{tikz}
\usepackage{authblk}
\usepackage[T1]{fontenc}
\usepackage{graphicx}
\usepackage{multirow}
\begin{document}
\title{Bridging the AI Adoption Gap: Designing an Interactive Pedagogical Agent for Higher Education Instructors
}
%
%

%
\author{
    Si Chen\inst{1} \and
    Reid Metoyer\inst{2} \and
    Khiem Le\inst{1} \and
    Adam Acunin\inst{1} \and
    Izzy Molnar\inst{1} \and
    Alex Ambrose\inst{1} \and
    James Lang\inst{1} \and
    Nitesh Chawla\inst{1} \and
    Ronald Metoyer\inst{1}
}
\authorrunning{S. Chen et al.}  

\institute{
    University of Notre Dame, Notre Dame, IN, USA \\
    \email{\{schen34,kle3,aacunin2,imolnar,gambrose,jlang4,nchawla,rmetoyer\}@nd.edu} 
    \and
    Columbia University, New York, NY, USA \\
    \email{ram2299@columbia.edu}
}
\maketitle              

\begin{abstract}Despite their pivotal role in integrating AI into education, instructors’ adoption of AI-powered tools remains inconsistent, with limited research on designing AI tools for broader adoption. This study employs a human-centered design and qualitative approach to investigate the design of interactive pedagogical agents that provide suggestions on instructors' questions. \textbf{First}, a formative study examined current pedagogical support strategies through interviews with five pedagogy experts. \textbf{Second}, a participatory design session engaged ten pedagogy experts in reviewing a storyboard illustrating chatbot features tailored to instructors with varying AI literacy and attitudes. Experts also evaluated the quality of LLM-generated suggestions based on instructors' common teaching challenges. Findings highlight the need for chatbot interactions that foster trust, particularly for AI-conservative instructors, by providing social transparency in peer usage and greater flexibility in how much or how little instructors engage with the system. Additionally, we propose design recommendations for enhancing AI-generated teaching suggestions, such as assessing prior teaching experiences. This work underscores the urgency of supporting AI-conservative instructors as AI literacy and attitudes intertwine, posing risks of pedagogical divides and diminished student learning opportunities if left unaddressed.


\end{abstract}
\section{Introduction}As AI continues to reshape education, its success depends not only on technical advancements but also on human-centered design that ensures AI tools align with instructors’ needs, expertise, and agency. While AI-powered tools have been widely explored for student learning—through automated assessment, adaptive learning platforms, and AI tutors \cite{ref_Cukurova} \cite{ref_Smolansky}—less attention has been given to how AI can support instructors in instructional design, assessment, and pedagogy.

Instructors play a crucial role in shaping AI adoption in classrooms, yet their engagement with AI tools varies widely. Research indicates that AI literacy and attitudes toward AI influence adoption: confidence in AI use increases the likelihood of adoption, while concerns about pedagogical autonomy, reliability, and ethical risks contribute to skepticism \cite{ref_Nourani} \cite{ref_long}. Institutional support, such as workshops offered by Centers for Teaching and Learning, has been identified as a key factor in facilitating AI adoption \cite{Ayanwale} \cite{ref_Nourani}. Without structured faculty support, instructors may hesitate to integrate AI into their teaching, potentially widening pedagogical divides—where some benefit from AI-enhanced instruction, while others remain disengaged \cite{ref_Wang}.  This disparity is particularly concerning in low-resourced contexts, where challenges related to digital access, teacher training, and infrastructure further hinder AI adoption \cite{Yadav}. Addressing this gap in teaching literacy is essential to make sure that AI adoption fosters equitable learning environments across various educational settings.


Advancements in large language models (LLMs) offer new opportunities for instructor-facing AI tools. LLMs, such as GPT-4 \cite{ref_gpt4}, generate contextually relevant responses, enabling interactive and adaptive pedagogical support \cite{ref_Kim} \cite{ref_lu}. While LLMs have been applied to automate some of the instructors' tasks, such as automated feedback \cite{Yan}, their role in supporting instructor decision-making remains underexplored. Unlike traditional AI systems that automate specific tasks, LLMs facilitate dynamic, real-time engagement, making them potentially valuable for instructors toward AI adoption. 

This study explores the design of interactive pedagogical agents, specifically how LLM-powered chatbots can generate teaching suggestions in response to instructors' questions. Pedagogical agents, also referred to as "learning partners," are characters displayed on a computer screen that guide users through multimedia learning environments \cite{Clarebout}. This use of ``agent'' differs from its meaning in AI, where agents are typically complex systems with planning capabilities \cite{Wang1}. Our research addresses the research question: \textit{How can interactive pedagogical agents be designed to promote broader AI adoption?}

\begin{figure}[!t]
    \centering
    \includegraphics[width=0.9\textwidth]{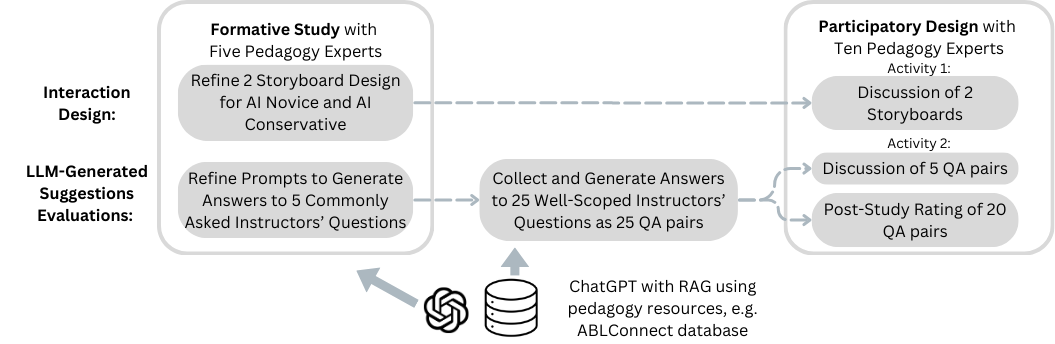}
    \caption{Two-Phase Study Design with Pedagogy Experts}
    \label{fig:study}
\end{figure}

To address this question, we conducted a two-phase study, as illustrated horizontally in Fig. \ref{fig:study}, with two objectives: (1) designing the interaction for suggestion-giving in response to instructors' questions, based on a set of example instructor questions and (2) evaluating LLMs' ability to generate teaching suggestions without additional training, to ultimately design a chatbot that can generate suggestions with a human-in-the-loop. Method-wise: First, a formative study explored pedagogy experts' current practices for providing teaching suggestions and working with instructors on AI adoption, drawing from their consulting experience. These insights informed the next phase, a participatory design study consisting of two activities: Activity 1- storyboarding with pedagogy experts to assess interaction designs for LLM-generated suggestions; Activity 2- expert evaluation of LLM-generated suggestions' quality, with recommendations for improvement. Our suggestion-generation approach integrates Retrieval-Augmented Generation (RAG) \cite{ref_rag1} to reduce hallucination risks by grounding LLM-generated suggestions in expert knowledge. Findings include interaction design principles to support AI adoption among AI-conservative instructors- low literacy and negative attitudes, including social transparency, greater flexibility in experimentation. 

\section{Related Work}
\subsection{AI in Education and Instructor Adoption}

AI adoption in education has primarily focused on student-facing applications, while instructor engagement remains limited \cite{ref_Cukurova} \cite{Ayanwale}. Key factors influencing adoption include AI literacy, pedagogical alignment, and institutional support. AI literacy and AI attitude are distinct yet interrelated—literacy refers to an instructor’s ability to evaluate and use AI, while attitude reflects an openness to AI adoption \cite{ref_Nourani} \cite{Ayanwale}. Confidence in teaching AI is a strong predictor of adoption, whereas AI anxiety has minimal impact \cite{Ayanwale}. The AI divide further exacerbates disparities, as instructors with lower AI literacy are less likely to engage with AI-driven teaching tools \cite{ref_Wang}.

Institutional policies also shape adoption. Many universities frame AI as a threat to academic integrity rather than a pedagogical asset, reinforcing skepticism \cite{Luo}. Additionally, student concerns about AI-instructor collaboration highlight issues, e.g., AI decontextualizing learning, bias, and shifting power dynamics \cite{Han}. User profiling studies suggest that individual beliefs about AI influence adoption behaviors, underscoring the need for tailored support \cite{ref_Nourani}.

Research on participatory design and instructor readiness suggests that actively involving educators in AI development fosters trust and increases adoption \cite{Lin} \cite{ref_Holstein}. Strategies that minimize workload, foster instructor ownership, and integrate AI into existing practices are critical for adoption. This study employs participatory design to support instructors with low AI literacy and negative attitudes—who are not yet considered in AI tool design.

\subsection{LLMs for Instructor Support}
Large language models (LLMs) provide new opportunities for instructor-facing AI tools, including automating instructional design, evaluating student work, and generating adaptive teaching recommendations \cite{ref_Hu}. However, research has primarily focused on student-facing applications, with limited exploration of LLMs as pedagogical support for instructors, particularly those with low AI literacy \cite{ref_Smolansky}. LLMs have been used to evaluate student-generated content \cite{Moore}, simulate student responses to aid question evaluation \cite{ref_lu}, and fine-tune chatbots for early math education \cite{Gattupalli}. These studies reveal LLM’s ability to support content development, but challenges remain in contextual alignment with instructor needs \cite{Zhang} \cite{ref_Hu}. Other studies user LLM as backbone for chat-based and quiz-based AI tutoring \cite{ref_Liu},

To support instructional decision-making, LLM-driven learning analytics dashboards provide insights into student-AI interactions, detect misuse, and assist in assessment refinement \cite{Pozdniakov} \cite{ref_kimm}. However, these systems often assume AI literacy and barely mention AI attitude, limiting their usefulness for a broad spectrum of instructors. Research on AI prompt engineering tools such as ChainForge \cite{Arawjo} and EvalLM \cite{ref_Kim} imply the importance of iterative refinement to enhance AI-generated outputs. Similarly, studies on intelligent tutoring systems emphasize the role of instructors in co-designing AI feedback mechanisms, ensuring alignment with pedagogical goals \cite{ref_Xia}. Other research on AI in assessment underscores the need for transparent, trust-building mechanisms \cite{ref_Smolansky}. 


This study advances prior research by designing an LLM-powered pedagogy agent in close collaboration with pedagogy experts—a domain expertise underexplored in LLM research. Often based in U.S. teaching centers, these experts drive instructional support, curriculum reform, and pedagogical innovation \cite{DeZure}, making them key partners in shaping AI-based teaching tools.

\section{Formative Study with Pedagogy Experts}

\subsection{Procedure}

This study gathered expert insights through group discussions on AI for teaching and learning, followed by individual clarifications. Five pedagogy experts in faculty development and instructional design, affiliated with different AI-related education product teams at the same institution, provided input on instructional support strategies, AI adoption barriers, and expectations for LLM-generated teaching suggestions. While they occasionally interacted, they did not collaborate directly. 
Key points were later clarified individually through follow-ups totaling approximately 60 minutes per expert, sometimes across multiple sessions. Sample questions included:  
\textit{- What have been the reasons instructors you work with like or dislike AI?  
- How easy or difficult is it to support them in AI adoption? Why? }  Five experts provided consent for note-taking and research usage. Thematic analysis \cite{them} of detailed notes identified patterns in instructional support, challenges in AI adoption, and expectations for LLM-generated suggestions. The research team conducted open-ended coding to refine these themes. 

\subsection{Themes on Design Insights} 

Analysis of the data revealed three key insights that shaped the next stage of research, where a storyboard prototype was designed and evaluated to explore differentiated interaction designs and suggestion delivery.  

\textbf{Common LLM tools did not accommodate end-users' AI literacy/attitude.}  
Existing AI-powered tools, such as ChatGPT, did not account for instructors’ varying levels of AI literacy. While some were comfortable experimenting with AI, others—particularly AI-conservatives—hesitated and relied on traditional teaching methods, such as paper-based materials and minimal use of learning management systems. Though universities offered AI-focused faculty workshops, participation required significant time and effort, limiting independent AI literacy growth. Ideally, AI tools would include built-in scaffolding to support instructors at different proficiency levels and foster gradual AI adoption.  

\textbf{Insight 1:} Two distinct storyboards were designed to explore interaction designs that accommodate a wide range of instructors: one for AI-conservative instructors, who have low AI literacy and a negative attitude, and another for AI-novice instructors, who have low AI literacy but a positive attitude. This classification does not suggest a strict dichotomy but elicits thinking from a spectrum of instructor perspectives. 

\textbf{F2F support is evidence-based but requires high commitment.}  
Traditional instructional support, such as consulting human teaching experts, reading pedagogical books, or joining faculty working groups, provided well-researched guidance. However, these methods demanded significant time and effort, making them inaccessible for instructors with tight schedules. Some also perceived seeking pedagogical support as signaling teaching weaknesses, further discouraging engagement. Despite these limitations, face-to-face (F2F) consultations followed a clear process: identifying challenges, providing examples, and refining suggestions, resulting in actionable teaching suggestions.  

\textbf{Insight 2:} The F2F consultation process was analyzed and incorporated into the chatbot interaction flow illustrated in the storyboards. The suggestion generation module in the next phase is based on existing pedagogy resources, specifically the ABLConnect database \cite{ablconnect}. ABLConnect is an online database of active learning efforts in college and university classrooms, curated by the Derek Bok Center for Teaching and Learning at Harvard.

\textbf{Experts should be involved in iteratively validating LLM suggestions.}  
Despite the growing use of AI in education, pedagogy experts were uncertain whether LLM-generated teaching suggestions would align with real-world classroom needs. Some were curious about the potential of LLM but unsure of its ability to generate contextually relevant and pedagogically sound suggestions. Validating LLM-generated content was necessary to determine its usefulness for instructional decision-making.

\textbf{Insight 3:} To evaluate current LLM’s potential, five common teaching questions and challenges were collected from pedagogy experts (see appendix). LLM-generated responses were then iteratively refined through multiple prompt adjustments and reviewed for usefulness in the next step.

\section{Participatory Design with Pedagogy Experts}
We conducted participatory design sessions with ten pedagogy experts experienced in faculty development and instructional design. Participants held roles such as learning designers, education researchers, teaching tech specialists, and teaching professors, with experience ranging from two to twenty years. One participant in this phase also took part in the previous session. Three described themselves as AI-conservatives, while the others were positive or neutral.

\subsection{Procedure}  
Sessions were conducted either remotely via Zoom or in person, each lasting approximately one hour. Each session began with an introduction to the chatbot’s purpose, followed by two activities, as illustrated in the right box of Fig. \ref{fig:study}. Activity 1 lasted 30 minutes, and Activity 2 took 15 minutes, followed by a post-study rating task that required an additional 20 minutes that participants completed on their own.

\begin{figure}[!t]
    \centering
    \includegraphics[width=0.6\textwidth]{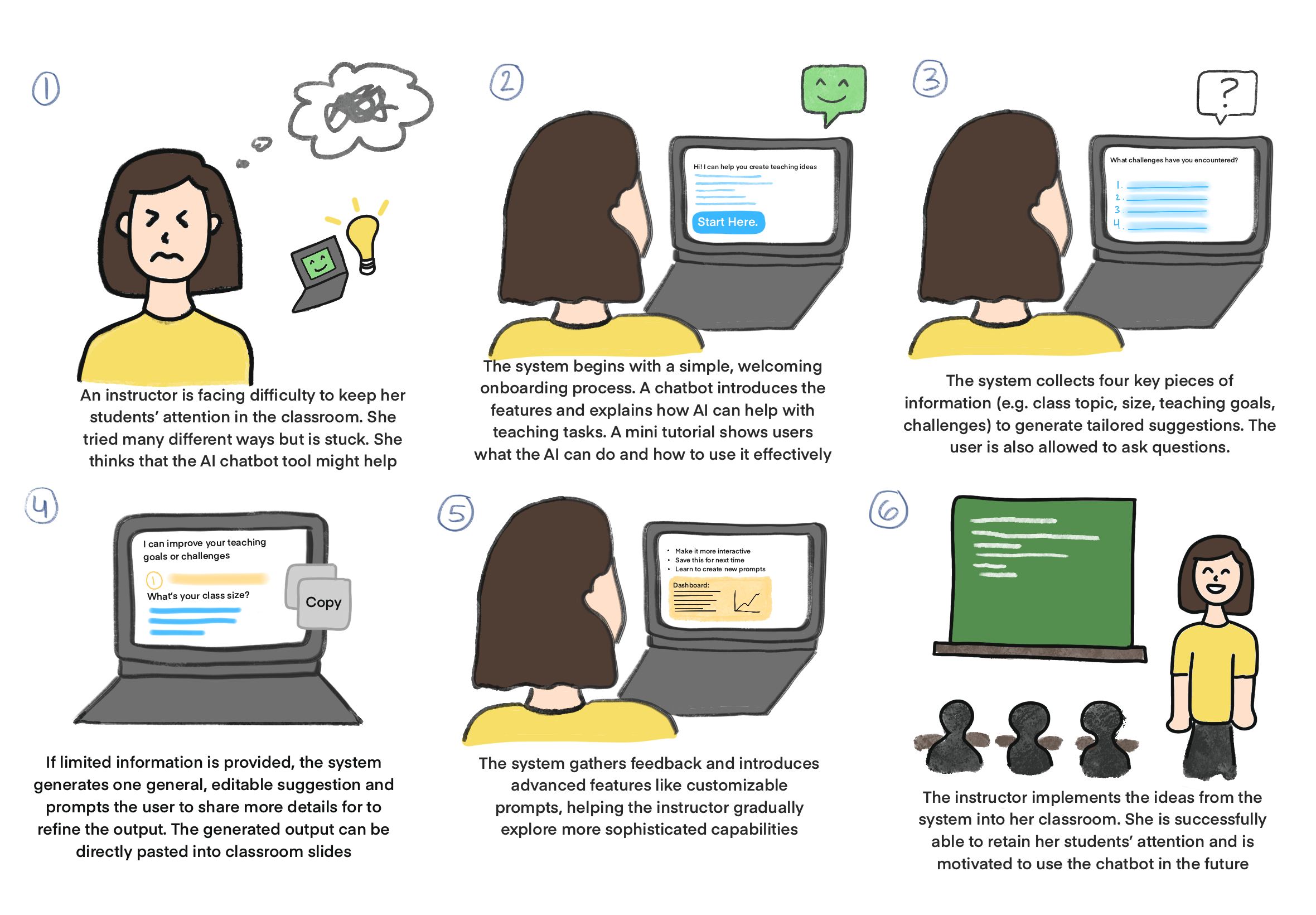}
    \caption{AI-Novice storyboard illustrating instructor-AI interaction.}
    \label{fig:storyboard2}
\end{figure}

\begin{figure}[!t]
    \centering
    \includegraphics[width=0.6\textwidth]{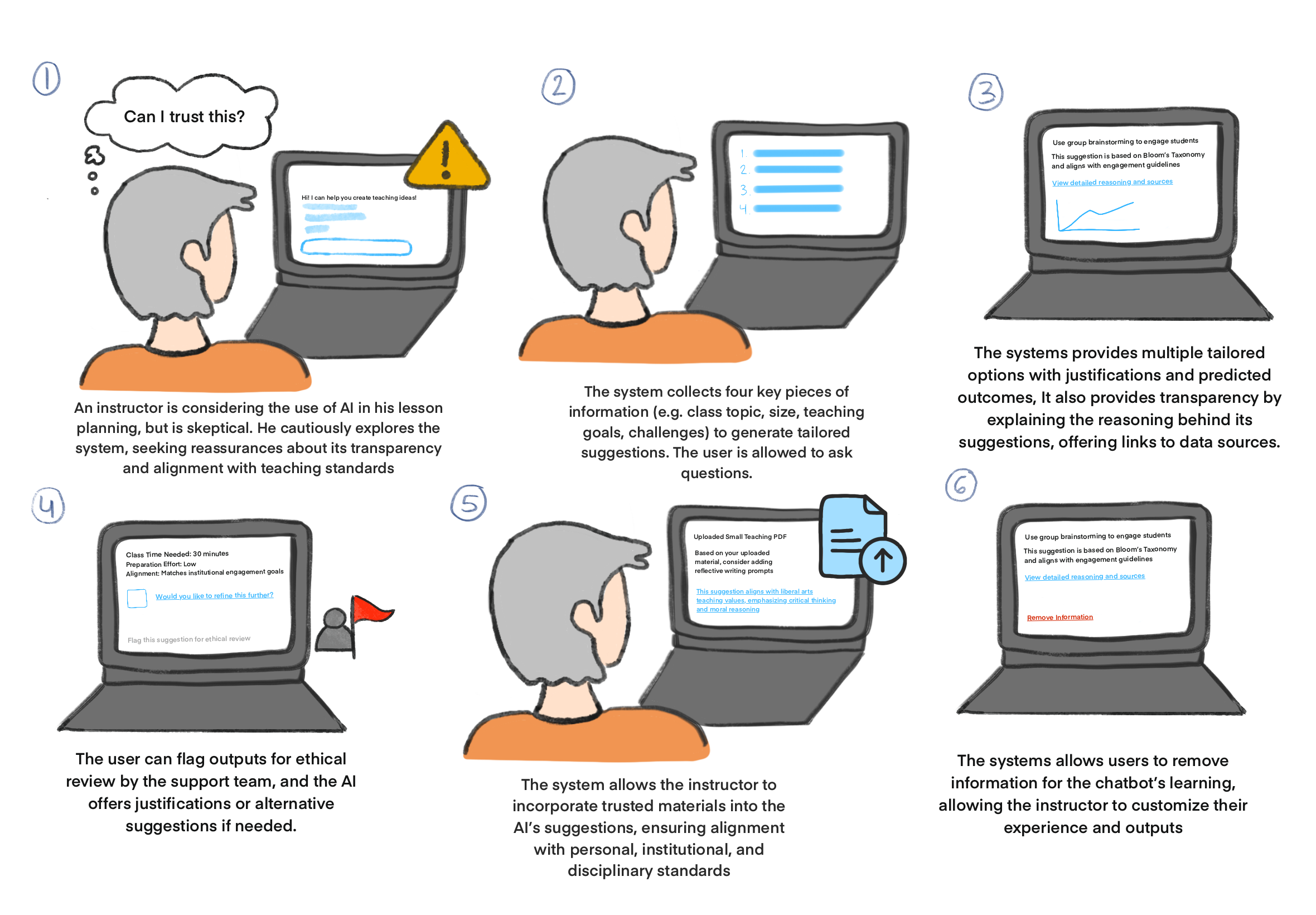}
    \caption{AI-Conservative Storyboard illustrating instructor-AI interaction.}
    \label{fig:storyboard1}
\end{figure}

\textbf{Activity 1: Storyboard Review} Storyboarding is a common technique in human-centered design for demonstrating system interfaces and contexts of use \cite{truong}. Participants reviewed two storyboards—one for AI-Novices in Fig. \ref{fig:storyboard2} and one for AI-conservative instructors in Fig. \ref{fig:storyboard1} —assessing user experience and feature usefulness. Their feedback included areas for refinement and how different instructors might engage with the tool.

\textbf{Activity 2: LLM-Generated Suggestion Evaluations} Following the storyboard review, experts evaluated LLM-generated teaching suggestions, assessing their relevance, accuracy, and alignment with effective teaching strategies. They compared different responses, selecting preferred outputs and discussing potential refinements. 
We further collected 25 commonly asked instructor questions that were well-scoped, as instructors sometimes pose overly broad questions (discussed in detail in the findings section). These questions were sourced from five pedagogy experts and used to acquire AI-assisted responses leveraging the database. The 25 QA pairs were then split into two sets: an initial set of 5 and a remaining set of 20. The first set of 5 QA pairs underwent an immediate qualitative evaluation by all ten pedagogy experts following the storyboard phase. The remaining 20 questions were assessed in a comparative evaluation, where pedagogy experts rated two different answers for each question—one generated by our prototype and one by default ChatGPT—resulting in a total of 40 responses to evaluate. Each response was rated across four dimensions using a 10-point scale, informed by pedagogy experts: Relevance (how well the suggestions match the user's intent), Actionability (how detailed and practical the suggestions are for implementation), Alignment (how well the suggestions align with effective teaching practices), and Reliability (the extent to which the responses can be trusted). Only five of the ten experts completed the rating due to time constraints.

\paragraph{Storyboard Design Details}  
Two storyboards were designed to explore different interactions: one tailored for instructors with limited AI literacy and another for AI-conservatives, both with low AI literacy and negative attitudes. These designs were informed by existing literature and insights from formative study sessions. These two versions allowed us to probe experts on how different attitudes toward AI influence engagement with the chatbot. The AI-Novice-Friendly Design prioritized simplicity, onboarding, and gradually introducing AI functionalities. In contrast, the AI-conservative Design emphasized transparency, offering detailed explanations, multiple response options, and alignment with institutional values. Before the formal session, storyboards underwent multiple rounds of expert feedback, refining key interaction flows and improving alignment with real-world F2F instructions flow, such as probing for more details.

\paragraph{Suggestion Generation Details}
The chatbot’s purpose is to aid instructors in leveraging the expertise of teaching and learning experts. It offers suggestions that users can include in their teaching to achieve the goals they present the tool with. To fulfill this requirement, we conducted six weekly brainstorming and assessment sessions with two pedagogy experts to determine how to best generate and format the output. Initially, we approached ChatGPT by giving it simple instructions to answer instructors’ five questions pooled from the experts and quickly determined that it had difficulty providing specific details about suggestions and was inconsistent with responses as it tended to give vague templates of activities to users. To address this challenge, we applied In-Context Learning \cite{ref_icl} and provided ChatGPT with examples written by teaching and learning experts for it to model after. 
Evaluation of these results suggested that the chatbot would require a more standardized database of examples to generate suggestions, ensuring that generated suggestions are grounded in expert knowledge. 

To this end, we first construct a standardized database of examples by leveraging available active learning suggestions from the ABLConnect database \cite{ablconnect}. After preprocessing and filtering irrelevant content, we end at a database of 115 high-quality examples. After that, with a user’s question, the three most similar examples to the question are retrieved from the database to guide the model’s response, following the concept of RAG \cite{ref_rag1}. This technique reduces hallucination risks, ensuring that LLM-generated suggestions are grounded in expert knowledge. Finally, a carefully designed instruction consisting of retrieved examples is constructed for feeding into the model, i.e., GPT-4, to generate suggestions.

\subsection{Analysis}  
The audio-recorded discussions on storyboarding and five QA pairs were transcribed and analyzed using thematic analysis \cite{them}. Two researchers, each with over a year of higher education teaching experience, independently reviewed the transcripts, identifying key themes related to chatbot features and LLM-generated suggestions. They then engaged in an discussions to refine codes and resolve discrepancies until a consensus was reached. A third researcher contributed only to theme discussions.  
Analysis began after the fourth expert session and continued iteratively until no new themes emerged, at which point recruitment was stopped, resulting in a total of ten participants. The post-study ratings of 20 QA pairs were averaged, and the mean scores are presented in Fig. \ref{rating}.

\subsection{Findings 1 - Designing Interactions for AI-Conservatives}

\begin{tcolorbox}[colframe=black,colback=gray!10,arc=2mm]
\textbf{Overview: }AI-Conservative Instructors avoid AI adoption from the beginning, reinforcing a cycle of low AI literacy and a negative attitude that further distances them from AI adoption. Interaction design should incorporate incremental information collection, scaffold AI adoption, and foster trust through social transparency. Encouraging instructors to refine LLM-generated suggestions through persona, tone design, and clear privacy assurances may help alleviate concerns.

\end{tcolorbox}
The analysis of storyboard data showcased the importance of considering AI-conservatives in interaction design to engage them with LLM-generated teaching suggestions effectively. See the appendix for a sample quote from each theme.

\paragraph{Easing AI Engagement Through Flexible and Incremental Information Collection}  
AI-Conservatives often hesitate to engage with AI because they feel pressured by how information is collected. Experts noted that AI-Conservatives would feel more comfortable and less overwhelmed if they could flexibly decide how much or how little information to provide. Instead of requiring all details upfront, the chatbot should adopt an incremental approach, allowing users to engage at their own pace. When less information is provided or when the initial question is broad, the chatbot could offer a list of general suggestions and let instructors choose how to proceed. This mirrors how experts conduct F2F consultations—rather than requiring instructors to fill out a detailed form; they guide them through options step by step. 
Therefore, it is important not to push for more information too soon. However, experts also emphasized that such a chatbot should not be designed to provide emotional support.

\paragraph{Scaffolding AI Adoption Through Progressive Changes in Teaching} 
Experts mention AI-Conservatives hesitate not only due to concerns about information disclosure but also because AI adoption challenges established teaching practices. Some of their hesitation stems not from this single chatbot but from broader fears and uncertainties about AI in teaching. Experts emphasized the importance of demonstrating the full range of AI capabilities while ranking them based on how much they alter instructors' practices and classroom time usage during panel 2 in the storyboard for AI-Novice. AI should gradually introduce features, starting with low-effort applications such as a 5-minute teaching tip while simultaneously exposing users to more advanced capabilities, such as generating discussion prompts, slides, podcasts, etc. The chatbot can encourage deeper engagement by progressively prompting users to explore advanced options during panel 4 in the storyboard for AI-Novice. For example, after suggesting a 5-minute Pair-and-Share activity, the chatbot could follow up with: "Would you like me to generate some questions for you?" This progressive approach makes AI adoption feel manageable rather than overwhelming. Tracking progress further supports hesitant users in the panel 5 storyboard for AI-Novice.  

\paragraph{Building Trust Through Peer Validation and Social Transparency}  
AI-Conservatives are more likely to explore AI-based tools when they see examples from colleagues rather than relying solely on generated suggestions. The chatbot should incorporate peer validation mechanisms, such as:  
- Testimonials from instructors who have successfully implemented similar strategies.  
- Recommendations from faculty within the same institution or discipline rather than generic generated suggestions.  
For instance, rather than stating, "Flipped classrooms improve engagement," the chatbot could say, "A faculty member in your department successfully used flipped learning in large lectures. Would you like to see their approach?" Expert 10 noted that AI-Conservatives are unlikely to experiment independently and require peer support to consider AI adoption from the beginning, posing panel 1 in the storyboard for AI-Conservatives to be less realistic.  

\paragraph{Encouraging Instructors' Refinement with Conversational Design}  
Experts emphasized that AI-Conservatives are more engaged when chatbot interactions feel like a collaborative discussion rather than a directive authority. The chatbot should present multiple options while allowing users to refine and adjust responses. Instead of offering rigid solutions, it should adopt a suggestive tone, such as, "Some instructors have found peer learning helpful. Would you like to explore this further?"  
To further engage users, the chatbot should allow suggestions to be challenged or adjusted rather than accepted as-is. A pop-up window, "Why this suggestion?" could explain the reasoning behind a response,  illustrated in panel 3 for AI-Conservative, and the chatbot could ask, "Would you like to see alternative approaches?" Expert 5 said that the flagging feature in panel 4 for AI-Conservative, helps enable users to critique generated responses. 

\paragraph{Addressing Data Privacy Concerns through Proactive and Reactive Strategies}  
Experts underscored that AI-Conservatives are cautious about sharing student data and instructional materials. To build trust, AI could adopt both proactive and reactive approaches. Proactively, the chatbot should clarify data handling policies before engagement, reassuring users with statements like, "Your information will not be stored or shared. You may also use anonymous examples."  Reactively, AI should offer on-demand explanations for generated responses and provide options to remove learned data, as illustrated in storyboard panel 6 for AI-Conservative.

\subsection{Findings 2 - Expert Evaluation of LLM-Generated Suggestions}

\begin{figure}[!h]
    \centering
    \includegraphics[width=0.65\textwidth]{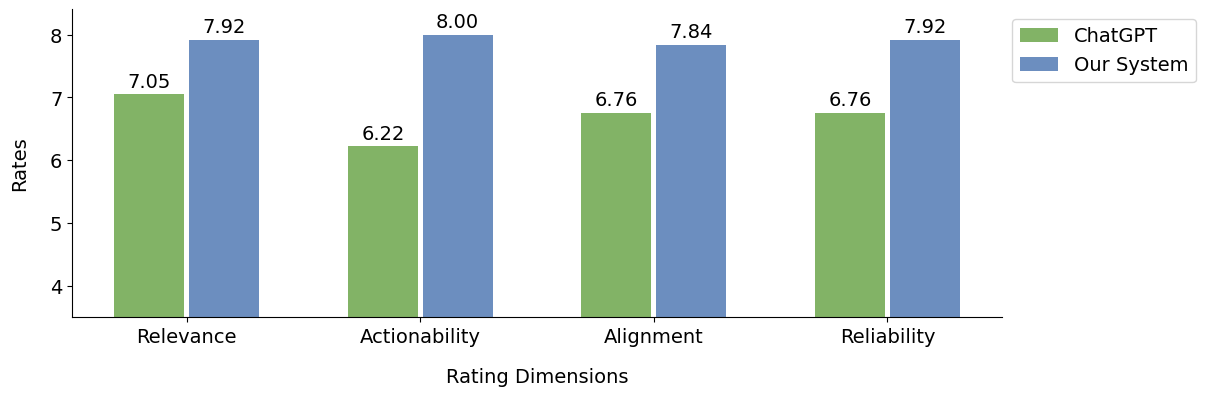}
    \caption{Averaged expert ratings for suggestions from our prototype and ChatGPT }
    \label{rating}
\end{figure}

Fig. \ref{rating} presents the averaged rating scores of five experts on the generated suggestions from our prototype compared to ChatGPT based on Relevance, Actionability, Alignment, and Reliability across 20 questions (Rated blindly, with the question sequence, answer, on 1-10 scale.). 

We draw observations from their evaluations. Experts consistently prefer our prototype’s suggestions over ChatGPT’s, rating them about one point higher across all dimensions and roughly two points on Actionability. Higher scores on Alignment and Reliability verify the effectiveness of integrating pedagogy resources, ensuring that generated suggestions are grounded in expert knowledge. Although demonstrated potential, the prototype still needs further improvement 

\paragraph{\textbf{Expert Feedback on Interactive Teaching Suggestions Generation}}
Experts outlined key steps for generating useful teaching suggestions through a chatbot that engages conversationally. Rather than collecting all details upfront and responding in a single turn, the chatbot should guide instructors through an interactive process. This process should refine queries, incorporate relevant context, and integrate past teaching experiences to generate suggestions.

\paragraph{Considerations for AI-Conservatives}
Experts noted that higher AI literacy enables instructors to confidently modify LLM-generated suggestions, recognizing that modifying prompts and providing more information can improve responses. These users are comfortable with ``good enough'' suggestions that they can adapt as needed.  In contrast, AI-Conservatives—often characterized by co-occurring negative AI attitudes and lower AI literacy—may feel overwhelmed by the perceived need for extensive input, leading to frustration and disengagement, as described before. The key approach is designing a conversational flow that effectively supports AI-conservative users, particularly in incremental information collection, that can generate good enough responses across the three steps below.

\paragraph{Step 1: Help Instructors Scope Teaching Questions/Goal}
Experts noted that vague or too specific teaching questions often led to generic or ineffective responses. The chatbot should assist instructors in refining their questions by prompting them to describe a specific recent teaching challenge they encountered.  One common example was \textit{inclusive teaching}, where instructors often lacked clarity on what they were asking. Inclusivity could relate to gender, race, assessment practices, or pre-class surveys, making it difficult and risky for LLM to generate a non-biased response. The chatbot should help instructors clarify their concerns/goals before providing suggestions.


\paragraph{Step 2: Collect Additional Context}
\textbf{Time Considerations}: Experts highlighted the need to assess the time and effort required to implement suggestions. Instructors sought estimates for both preparation time and in-class execution to evaluate feasibility. The chatbot should clearly provide this information for each recommendation or ask instructors their expectation prior to generation.  \textbf{Class Context}: Suggestions should account for practical constraints, as engagement tricks for large lecture courses differ from those for small discussion-based seminars. Additional factors, such as student grade level and whether a course is required or elective, also influence suggestion usefulness.

\paragraph{Step 3: Evaluate Instructors' Actual Teaching Ability}
Experts recommended that the chatbot collect relevant background details about an instructor’s teaching experience to tailor suggestions more effectively. These considerations included:  Years of teaching experience;  How they initially learned to teach (e.g., formal training, self-guided learning, mentorship);  Whether they had engaged with pedagogical literature,  Participation in teaching workshops or AI training programs.  Experts cautioned that teaching experience doesn't always equate to instructional expertise. Even long-time instructors may be unfamiliar with pedagogical strategies, resistant to change due to their established methods, or teaching a new class. Therefore, generated suggestions shouldn't assume expertise based solely on years of experience.

\paragraph{Typical Errors: Suggestion Fusion and Lack of Discipline Specificity}  
Experts noted that the generated responses sometimes combined unrelated suggestions without a clear rationale, resulting in fragmented or impractical suggestions that could confuse students. While blending teaching suggestions is not inherently wrong, experts emphasized that they typically begin with a single suggestion, ask instructors to implement it to meet specific learning goals, and adapt afterward. The suggestion should also be based on pedagogical research or a curated list of evidence-based suggestions. Ensuring alignment with discipline-specific teaching methods, particularly in STEM education, would further improve usefulness.

\subsection{Future Works on Improving Suggestion Generation} 

\textbf{Data Collection:} A key challenge in developing this interactive generation system is the lack of diverse instructor-AI conversational data, limiting its ability to model real-world teaching contexts and AI literacy levels. To address this, expert feedback will inform the development of an instructor simulator that models varying AI literacy levels and attitudes. The simulator will engage experts in structured interactions to collect context-rich suggestion-giving data, particularly for AI-conservative instructors. Additionally, current LLM-generated suggestions rely on participant-submitted questions and the ABLConnect database \cite{ablconnect}, limiting diversity. Future work should incorporate multiple reliable pedagogy resources. \textbf{Model Fine-tuning:} Expert-rated responses in this study could be used to refine AI using reinforcement learning from human preferences \cite{ref_rlhf}.  While RAG \cite{ref_rag1} performed well, more advanced techniques should be explored to refine LLM reasoning further.   \textbf{Instructor Involvement:} AI literacy and attitude impact engagement. Future work should incorporate instructor interactions, not just expert-driven refinements. Instructors should actively refine AI-generated suggestions to ensure adaptability across varying teaching styles and experiences.
\section{Conclusion}
This study explored the design of an LLM-powered interactive pedagogical agent to generate teaching suggestions for instructors with limited AI literacy. Findings highlight the importance of social transparency and incremental information collection in instructor-AI interaction design to foster AI adoption, particularly among AI-conservative instructors.  Through a human-centered approach, this work demonstrates how pedagogical agents can help bridge gaps in faculty development Future research should investigate how human-in-the-loop AI tools can further refine interaction design and optimize LLM-generated suggestions to better support a wide range of instructors.



\appendix  
\section{Appendix}
\begin{itemize}
    \item Table \ref{tab:expert_quotations_1}: Storyboard Themes Quotations
    \item Table \ref{tab:expert_quotations_2}: LLM-Generated Suggestions Themes Quotations
    \item Table \ref{tab:questions}: Questions provided by experts for evaluating generated suggestions.
    \item Table \ref{example-chatgpt}: An example of generated suggestions from ChatGPT.
    \item Table \ref{example-ours}: An example of generated suggestions from our prototype.
\end{itemize}

\clearpage
\begin{table}
    \centering
    \begin{tabular}{|p{3cm}|p{9cm}|l|}
        \hline
        \textbf{Theme} & \textbf{Quotation} & \textbf{Expert} \\ \hline
        \raggedright Easing AI Engagement Through Flexible and Incremental Information Collection & \raggedright “[The] first time [the] chatbot may need to know more about the instructor... their experiences, preferences, but the second time maybe [the] chatbot will ask a question, like some reflection questions [about] what's going wrong. For a specific course, they can use the same chain or thread. Because… that chain includes all the relevant info about that course.” & 1 \\ \hline
        \raggedright Scaffolding AI Adoption Through Progressive Engagement & \raggedright “That would be a great way for you to track it if they had a reflection piece within there to where they could go when they come back and then say, how did it go? And you’re like… this didn't work out or whatever else like that.” & 4 \\ \hline
        \raggedright Building Trust Through Peer Validation and Social Transparency & \raggedright “Maybe they… meet some challenges… when they teach their courses and they read an article about [how] their peers… solve this problem [using the chatbot]… I think it should be someone from the *** community that they feel relevant to… build a connection between the community then they will be more likely to try this tool.” & 1 \\ \hline
        \raggedright Encouraging Instructors' Refinement with Conversational Design & \raggedright “I certainly like the idea here of the tool, especially for the skeptic, although for what it's worth, I actually think there's value in this across the board, even for the kind of more novice or optimistic user of having the tool be transparent about where it's drawing its information from or what its line of reasoning is… All of those… features of transparency that the instructor can follow along and interrogate and try to trace back sources of information that’s pulling from and then give them the ability to flag areas of concern or adjust the information that the tool has like I think those are good models for these AI tools in general” & 5 \\ \hline
        \raggedright Addressing Data Privacy Concerns through Proactive and Reactive Strategies & \raggedright “Open the opportunity for them to challenge AI, [it] can be… a good way to ease their concern... Since you are using AI chat bot maybe give them an opportunity at the beginning and say, okay, do you have any ethical concern[s] feel free to ask me or have some guiding question for them to ask AI or follow-up” & 3 \\ \hline
    \end{tabular}
    \caption{Storyboard Themes Quotations}
    \label{tab:expert_quotations_1}
\end{table}

\newpage

\begin{table}
    \centering
    \begin{tabular}{|p{3cm}|p{9cm}|l|}
        \hline
        \textbf{Theme} & \textbf{Quotation} & \textbf{Expert} \\ \hline
        \raggedright Considerations for AI-Conservatives & \raggedright “When I'm reading this, I'm thinking what nudges the instructor to use the AI tool because he is very cautious. I'm thinking that he might be cautious about privacy and his course content... if the instructors are skeptical and very cautious they won't try the tool and they won't even upload the course content, course materials, even though they are told that this information will be removed.” & 1 \\ \hline
        \raggedright Step 1: Help Instructors Scope Teaching Questions/Goal & \raggedright “Yeah, [the prompt is] just too broad, so [you need] to get a conversation: Are there specific aspects of your teaching you would want to be including? For example, in your assignments, classroom discussions, classroom community and you can say like inclusive teaching covers all these different things, is there a specific area that you would want to focus on?… Because otherwise, a whole list of things is sort of overwhelming.” & 8 \\ \hline
        \raggedright Step 2: Collect Additional Context & \raggedright “We should also prompt them for… how much effort they’re hoping to spend… like how much time are you willing to invest ahead of time and how much of your class time are you able to spend on this?” & 7 \\ \hline
        \raggedright Step 3: Evaluation of Instructors’ Actual Teaching Ability & \raggedright “I think the chatbot needs to understand what levels, not just what class topic. Is this a hundred level with first year students? Is it a four hundred level with seniors? Is it a major course? Is it an elective course? These are all major factors that will influence pedagogy. Also maybe asking how experienced they are [at] teaching. The chatbot needs to know, is this their first year teaching, are they a veteran, an older instructor that’s trying to modernize a traditional method… Is this a brand new course? Or is this a course [they have] taught many times before?” & 10 \\ \hline
        \raggedright Typical Errors: Suggestion Fusion and Lack of Discipline Specificity & \raggedright “I’m just guessing that [the chatbot] shoved some activities together in a way that doesn't actually play out super hopefully. [Mismatching teaching suggestion] can be [a good idea], but it needs vetting.” & 7 \\ \hline
    \end{tabular}
    \caption{LLM-Generated Suggestions Themes Quotations}
    \label{tab:expert_quotations_2}
\end{table}

\begin{table}
\centering
\begin{tabular}{|l|p{15cm}|}
\hline
\textbf{Set}                  & \textbf{Question}                                                                                                                                                                                                                                                                                     \\ \hline
\multirow[t]{5}{*}{5}   & Please give me some ideas of activities to do in class that reinforce the key concepts from a homework assignment but in a way that engages students and has them learning actively.                                                                                                         \\ \cline{2-2} 
                     & What are some scenarios I could present to students in order to have them apply their learning of key course concepts?                                                                                                                                                                       \\ \cline{2-2} 
                     & What are some ways I could involve students more in the creation of course materials like the syllabus and course readings so that they have a greater sense of investment in the course?                                                                                                    \\ \cline{2-2} 
                     & What are some ways I could get the class to create community guidelines for the course so that students have a greater sense of investment in and responsibility toward one another?                                                                                                         \\ \cline{2-2} 
                     & What are some ways I can shape or phrase course policies so that students understand I am invested in their learning and not interested in policing them?                                                                                                                                    \\ \hline
\multirow[t]{20}{*}{20} & What are some writing assignment prompts that encourage students to invest themselves authentically in the process of writing?                                                                                                                                                               \\ \cline{2-2} 
                     & What are some multimodal assignments I could assign students that would allow them to use generative AI to assist in their work but wouldn't circumvent the necessary cognitive and intellectual work they need to do?                                                                       \\ \cline{2-2} 
                     & What are some assignments I could develop that have real-world stakes and impacts for students, so that they feel more palpably that they are writing for a purpose?                                                                                                                         \\ \cline{2-2} 
                     & What are some of the ways I might scaffold an assignment across several weeks so that students engage their work over time and develop a process?                                                                                                                                            \\ \cline{2-2} 
                     & What are some ways I might design an assignment so that students feel comfortable taking risks and aren't concerned about being penalized for failure, even though they eventually have to get a grade on the assignment?                                                                    \\ \cline{2-2} 
                     & What are some in-class activities I could do in the classroom to create a greater sense of belonging for everyone in the room?                                                                                                                                                               \\ \cline{2-2} 
                     & How might I use Canvas/my LMS to create a greater sense of belonging and community in my course?                                                                                                                                                                                             \\ \cline{2-2} 
                     & What are some design strategies and techniques I could implement to make my course materials (like my syllabus and assignment prompts) more welcoming and inclusive?                                                                                                                         \\ \cline{2-2} 
                     & I am looking for some potential assessment methods that will capture the learning growth of each student and that meet the student where they're coming from rather than assess them against a shared evaluative standard.                                                                   \\ \cline{2-2} 
                     & What are some assessment practices that would allow me to better capture the learning students undergo during the process of creating their work rather than evaluating the outcome of that process?                                                                                         \\ \cline{2-2} 
                     & What are some assignments or assessment practices that encourage greater metacognitive reflection in my students toward their learning?                                                                                                                                                      \\ \cline{2-2} 
                     & I'm teaching a large introductory course and finding it challenging to ensure all my learning activities, both in-class and out-of-class, effectively support the course's learning objectives. Could you suggest ways to better align my learning content with course goals?                \\ \cline{2-2} 
                     & I feel overwhelmed by the time it takes to design engaging and rigorous lessons while managing grading, research, and meetings. How can I balance my workload and maintain intellectual rigor when designing lessons?                                                                        \\ \cline{2-2} 
                     & I'm exploring new ways to use technology to enhance student engagement inside and outside the classroom. Could you provide suggestions for leveraging technology to create active learning experiences and promote student interaction?                                                      \\ \cline{2-2} 
                     & I teach a course with a diverse student population. Some students have extensive background knowledge, while others are new to the subject. How can I pace my instruction and design activities to ensure all students are both challenged and supported in achieving the learning outcomes? \\ \cline{2-2} 
                     & Should I share rubrics with students before assessments are due?                                                                                                                                                                                                                             \\ \cline{2-2} 
                     & What is the best method for exam assistance in STEM courses: a provided equation sheet, let the students write a “cheat sheet” or open book?                                                                                                                                                 \\ \cline{2-2} 
                     & How do I encourage attendance without requiring it as part of the grade? Are there different ways of “convincing” students to come to class?                                                                                                                                                 \\ \cline{2-2} 
                     & What are some good activities/tricks for learning and remembering student names in large classrooms?                                                                                                                                                                                         \\ \cline{2-2} 
                     & I need students to memorize a list of plants’ scientific names. Can you suggest classroom techniques to help them do this?                                                                                                                                                                   \\ \hline
\end{tabular}
\caption{Questions provided by experts for evaluating generated suggestions.}
\label{tab:questions}
\end{table}

\begin{table}
\begin{tabular}{|l|p{14cm}|}
\hline
Question      & What are some scenarios I could present to students in order to have them apply their learning of key course concepts? \\ \hline
Suggestions & 1. **Case Studies**: Present real-world scenarios related to the course content. Ask students to analyze the situation, identify key concepts, and propose solutions or outcomes based on their learning.\newline 2. **Role-Playing**: Create scenarios where students must assume roles relevant to the course material. This encourages them to apply concepts in a dynamic and interactive way, fostering deeper understanding.\newline 3. **Problem-Based Learning**: Design complex, open-ended problems that require students to apply multiple course concepts to find a solution. This approach encourages critical thinking and collaboration.\newline 4. **Simulations**: Use simulations to mimic real-life processes or systems. This allows students to experiment with course concepts in a controlled environment, observing outcomes and refining their understanding.\newline 5. **Debates**: Organize debates on topics related to the course. This requires students to research, apply their knowledge, and articulate their understanding of key concepts while considering opposing viewpoints.\newline 6. **Project-Based Learning**: Assign projects that require students to apply course concepts to create something tangible, such as a report, presentation, or model. This approach integrates learning with practical application.\newline 7. **Scenario-Based Discussions**: Present hypothetical scenarios and facilitate discussions where students must apply their knowledge to predict outcomes or solve problems, encouraging them to think critically and collaboratively.\newline 8. **Interactive Quizzes**: Use technology to create interactive quizzes that present scenarios requiring application of course concepts. This can be a fun and engaging way to reinforce learning.\newline 9. **Peer Teaching**: Have students create and present scenarios to their peers, explaining how they would apply course concepts. This reinforces their understanding and builds communication skills.\newline 10. **Reflective Journals**: Ask students to write about how they would apply course concepts to hypothetical or real-life scenarios in a journal. This encourages reflection and deeper understanding of the material. \\ \hline
\end{tabular}
\caption{An example of generated suggestions from ChatGPT.}
\label{example-chatgpt}
\end{table}

\begin{table}
\begin{tabular}{|l|p{14cm}|}
\hline
Question      & What are some scenarios I could present to students in order to have them apply their learning of key course concepts? \\ \hline
Suggestions & **Scenario-Based Learning Suggestions**\newline **Purpose:** To provide students with practical scenarios that allow them to apply key course concepts in a realistic context, enhancing their understanding and retention of the material.\newline **Introduction and Background:**  \newline Scenario-based learning is an effective educational strategy that immerses students in real-world situations where they must apply their knowledge and skills. This approach not only reinforces theoretical concepts but also develops critical thinking, problem-solving, and decision-making abilities. By engaging with scenarios, students can see the relevance of their learning and are better prepared for real-life applications.\newline **Procedure:**\newline **1. Scenario Development:**\newline - **Identify Key Concepts:** Begin by identifying the core concepts or skills you want students to apply. These should align with your course objectives.\newline - **Create Realistic Scenarios:** Develop scenarios that are relevant to the students' field of study or future career paths. Ensure these scenarios are complex enough to challenge students but not so difficult that they become discouraged.\newline - **Incorporate Multiple Perspectives:** Design scenarios that require students to consider different viewpoints or roles, encouraging them to think critically and empathetically.\newline **2. Scenario Presentation:**\newline - **Introduce the Scenario:** Present the scenario to the students, providing all necessary background information. This can be done through written descriptions, videos, or role-playing activities.\newline - **Clarify Objectives:** Clearly outline what students are expected to achieve by the end of the scenario exercise. This could include identifying problems, proposing solutions, or making decisions based on the information provided.\newline **3. Scenario Engagement:**\newline - **Group Work:** Divide students into small groups to encourage collaboration and discussion. Each group can work on the same scenario or different ones, depending on the class size and objectives.\newline - **Guided Exploration:** Provide guiding questions or prompts to help students navigate the scenario. Encourage them to apply course concepts and consider the implications of their decisions.\newline - **Facilitate Discussion:** As students work through the scenario, circulate the room to offer support, answer questions, and prompt deeper thinking.\newline **4. Scenario Debrief:**\newline - **Group Presentations:** Have each group present their findings and solutions to the class. This allows students to learn from each other's perspectives and approaches.\newline - **Class Discussion:** Facilitate a class-wide discussion to reflect on the different solutions proposed, the application of course concepts, and any challenges encountered during the exercise.\newline - **Feedback and Reflection:** Provide feedback on the students' application of concepts and encourage them to reflect on what they learned and how they might apply it in the future.\newline **Follow-up:**\newline - **Written Reflection:** Ask students to write a brief reflection on the scenario exercise, focusing on what they learned and how they applied course concepts.\newline - **Scenario Revisions:** Based on student feedback and performance, revise scenarios for future use to better align with learning objectives and student needs.\newline **Comments:** Scenario-based learning is a dynamic way to engage students and deepen their understanding of course material. By simulating real-world situations, students can see the practical application of their learning, which enhances motivation and retention. \\ \hline
\end{tabular}
\caption{An example of generated suggestions from our prototype.}
\label{example-ours}
\end{table}

\end{document}